\RequirePackage[mathlines]{lineno} 
\documentclass[prl,twocolumn,showpacs,amsmath,amssymb]{revtex4}
\usepackage{overpic,graphicx}
\usepackage{dcolumn}
\usepackage{bm}
\usepackage{rotating}
\usepackage{subfigure}
\usepackage{color}

\renewcommand{\thetable}{\arabic{table}}

\begin{document}

\title{\bf \boldmath
Measurement of
the Branching Fraction For the Semi-Leptonic Decays $D^{0(+)}\to \pi^{-(0)}\mu^+\nu_\mu$ and Test of Lepton Flavor Universality
}

\author{M.~Ablikim$^{1}$, M.~N.~Achasov$^{9,d}$, S. ~Ahmed$^{14}$, M.~Albrecht$^{4}$, A.~Amoroso$^{53A,53C}$, F.~F.~An$^{1}$, Q.~An$^{50,40}$, J.~Z.~Bai$^{1}$, Y.~Bai$^{39}$, O.~Bakina$^{24}$
% ...
\\
\vspace{0.2cm}
(BESIII Collaboration)\\
\vspace{0.2cm} {\it
$^{1}$ Institute of High Energy Physics, Beijing 100049, People's Republic of China\\
% ...
$^{4}$ Bochum Ruhr-University, D-44780 Bochum, Germany\\
% ...
$^{9}$ G.I. Budker Institute of Nuclear Physics SB RAS (BINP), Novosibirsk 630090, Russia\\
% ...
$^{14}$ Helmholtz Institute Mainz, Johann-Joachim-Becher-Weg 45, D-55099 Mainz, Germany\\
% ...
$^{24}$ Joint Institute for Nuclear Research, 141980 Dubna, Moscow region, Russia\\
% ...
$^{39}$ Southeast University, Nanjing 211100, People's Republic of China\\
$^{40}$ State Key Laboratory of Particle Detection and Electronics, Beijing 100049, Hefei 230026, People's Republic of China\\
% ...
$^{50}$ University of Science and Technology of China, Hefei 230026, People's Republic of China\\
% ...
$^{53}$ (A)University of Turin, I-10125, Turin, Italy; (B)University of Eastern Piedmont, I-15121, Alessandria, Italy; (C)INFN, I-10125, Turin, Italy\\
% ...
$^{d}$ Also at the Novosibirsk State University, Novosibirsk, 630090, Russia\\
% ...
}
}

\begin{abstract}
Using a data sample corresponding to an integrated luminosity of $2.93\,\rm fb^{-1}$ taken at a center-of-mass energy of 3.773\,GeV with the BESIII detector operated at the BEPCII collider, we perform an analysis of the semi-leptonic decays $D^{0(+)}\to \pi^{-(0)}\mu^+\nu_\mu$.
The branching fractions of $D^0\to \pi^-\mu^+\nu_\mu$ and $D^+\to \pi^0\mu^+\nu_\mu$ are measured to be $(0.272 \pm 0.008_{\rm stat.} \pm 0.006_{\rm syst.})\%$ and $(0.350 \pm 0.011_{\rm stat.} \pm 0.010_{\rm syst.})\%$, respectively, where the former is of much improved precision compared to previous results and the latter is determined for the first time. Using these results along with previous BESIII measurements of $D^{0(+)}\to \pi^{-(0)}e^+\nu_e$, we calculate the branching fraction ratios to be ${\mathcal R}^0\equiv   {\mathcal B}_{D^{0}\to \pi^{-}\mu^+\nu_\mu}/{\mathcal B}_{D^{0}\to \pi^{-}e^+\nu_e}=0.922\pm 0.030_{\rm stat.}\pm0.022_{\rm syst.}$ and ${\mathcal R}^+\equiv {\mathcal B}_{D^{+}\to \pi^{0}\mu^+\nu_\mu}/{\mathcal B}_{D^{+}\to \pi^{0}e^+\nu_e}=0.964\pm 0.037_{\rm stat.}\pm0.026_{\rm syst.}$, which are compatible with the theoretical expectation of lepton flavor universality within $1.7\sigma$ and $0.5\sigma$, respectively. We also examine the branching fraction ratios in different four-momentum transfer square regions, and find no
significant deviations from the standard model predictions.
\end{abstract}


\maketitle

In the standard model (SM),
the couplings of leptons to gauge bosons are expected to be independent of lepton flavors. This property is known as lepton flavor universality~(LFU)~\cite{Xing2012,Fajfer2015,BFajfer2012,Fajfer2012,Bauer2016}.
Tests of LFU with
semileptonic~(SL) decays of pseudoscalar mesons provide powerful probes of new physics beyond the SM.
In recent years, BaBar, Belle and LHCb experiments reported tests of LFU in various SL $B$ decays.
The measured branching fraction (BF) ratios
${\mathcal B}_{B\to \bar D^{(*)}\tau^+\nu_\tau}/{\mathcal B}_{B\to \bar D^{(*)}\ell^+\nu_\ell}$~($\ell=\mu$, $e$)
~\cite{babar_1,babar_2,belle_1,belle_3,refereeb2,lhcb_1}
and
${\mathcal B}_{B\to K^{(*)}\mu^+\mu^-}/{\mathcal B}_{B\to K^{(*)}e^+e^-}$~\cite{lhcb_kee_1,JHEP_refereeA} 
deviate from the SM predictions by 1.6-2.7 and 2.1-2.6 standard deviations, respectively.
In view of this, tests of LFU in the charm sector using the SL $D$ decays are important complementary tests.

This Letter presents tests of LFU in $D^{0(+)}\to \pi^{-(0)}\ell^+\nu_\ell$ decays~\cite{charge} at BESIII.
Recently, the Cabibbo-favored decays $D^{0(+)}\to\bar K\ell^+\nu_\ell$
were precisely studied at BESIII, and the measured BF ratios (BFRs)
${\mathcal B}_{D \to \bar K\mu^+\nu_\mu}/{\mathcal B}_{D\to \bar K e^+\nu_e}$
are compatible with the SM expectations~\cite{bes3_piev,bes3_pi0ev,bes3_k0barmuv,bes3_kmuv}.
Nevertheless, tension between previous measurement and the SM prediction
for the Cabibbo-suppressed decays $D^{0}\to \pi^-\ell^+\nu_\ell$ is found.
In the SM,
the BFRs ${\mathcal R}^{0(+)}_{\rm LFU}={\mathcal B}_{D^{0(+)}\to \pi^{-(0)}\mu^+\nu_\mu}/{\mathcal B}_{D^{0(+)}\to \pi^{-(0)}e^+\nu_e}$ are expected to be $0.985\pm0.002$~\cite{arxiv0985},
which deviates from unity due to different phase space available to the two processes.
With the world-average values of ${\mathcal B}_{D^{0}\to \pi^{-}\mu^+\nu_\mu}$ and ${\mathcal B}_{D^{0}\to \pi^{-}e^+\nu_e}$~\cite{pdg2014},
${\mathcal R}^0_{\rm LFU}$ is 17\% lower than the SM prediction, corresponding to $2.1$ standard deviations.
Currently, the most precise measurements of
${\mathcal B}_{D^{0(+)}\to \pi^{-(0)}e^+\nu_e}$ have reached an accuracy
better than 3\%~\cite{bes3_piev,bes3_pi0ev}.
However, the world-average value of ${\mathcal B}_{D^{0}\to \pi^{-}\mu^+\nu_\mu}$
has a large relative uncertainty of 10\%~\cite{pdg2014,belle_pimunu,focus_pimunu},
and the decay $D^{+}\to \pi^{0}\mu^+\nu_\mu$ has not been measured.
To clarify this tension, it is crucial to precisely measure ${\mathcal B}_{D^{0(+)}\to \pi^{-(0)}\mu^+\nu_\mu}$.

The analysis is performed
by using a data sample corresponding to an integrated luminosity of $2.93\,\rm fb^{-1}$~\cite{BESIII292} taken at
a center-of-mass energy of $\sqrt s= 3.773$\,GeV with the BESIII detector.
Details about the design and performance
of the BESIII detector are given in Ref.~\cite{BESIII}.
A {\sc geant4}-based~\cite{geant4} Monte Carlo (MC) simulation software package, which includes a description of the detector geometry and its response, is used to determine
the detection efficiency and to estimate potential backgrounds.
An `inclusive' MC sample corresponding to about 10 times the luminosity of data is produced at $\sqrt s=3.773$\,GeV. It includes
the $D^0\bar D^0$, $D^+D^-$, and non-$D\bar D$ decays of $\psi(3770)$,
the initial state radiation (ISR) production of $\psi(3686)$ and $J/\psi$, and
the $q\bar q$ ($q=u$, $d$, $s$) continuum process, along with
Bhabha scattering, $\mu^{+}\mu^{-}$ and $\tau^{+}\tau^{-}$ events.
The production of $\psi(3770)$ is simulated by the MC generator {\sc kkmc}~\cite{kkmc}.
The measured decay modes of the charmoniums
are generated using {\sc EvtGen}~\cite{evtgen} with the BFs reported in Ref.~\cite{pdg2012},
and the remaining decay modes are generated using {\sc LundCharm}~\cite{lundcharm}.
The signal $D^{0(+)}\to\pi^{-(0)}\mu^+\nu_\mu$ decays are simulated incorporating the
modified pole model~\cite{BK-model},
where the parameters of vector and scalar hadronic form factors (HFFs) are taken from
Refs.~\cite{bes3_piev,bes3_pi0ev,haiyang}.
The ISR effects~\cite{isr} and final state radiation (FSR) effects of all particles~\cite{photons} have
been included in the event generation.

At $\sqrt{s}=3.773$\,GeV, the $\psi(3770)$ resonance decays mainly into a $D\bar{D}$ pair. Throughout the text, $D$ refers to $D^0(D^+)$ and $\bar{D}$ refers to $\bar{D}^0(D^-)$ unless stated explicitly. If a $\bar{D}$ meson [called single-tag (ST)
$\bar{D}$ meson] is fully reconstructed,
the presence of a $D$ meson is guaranteed.
Thus, in the system recoiling against a ST $\bar{D}$ meson,
the SL decay $D^{0(+)}\to \pi^{-(0)}\mu^+\nu_\mu$
[called double-tag (DT) event] can be selected.
In this analysis, the ST $\bar{D}^0$ mesons are reconstructed using three hadronic decay modes:
$K^+\pi^-$, $K^+\pi^-\pi^0$ and $K^+\pi^-\pi^-\pi^+$,
while the ST $D^-$ mesons are reconstructed using six hadronic decay modes: $K^{+}\pi^{-}\pi^{-}$,
$K^0_{S}\pi^{-}$, $K^{+}\pi^{-}\pi^{-}\pi^{0}$, $K^0_{S}\pi^{-}\pi^{0}$, $K^0_{S}\pi^{+}\pi^{-}\pi^{-}$
and $K^{+}K^{-}\pi^{-}$. The BF of $D^{0(+)}\to \pi^{-(0)}\mu^+\nu_\mu$ is determined according to
\begin{equation}
{\mathcal B}_{D^{0(+)}\to \pi^{-(0)}\mu^+\nu_\mu} = N^{0(+)}_{\rm DT}/(N^{0(+)}_{\rm ST}\epsilon^{0(+)}_{\pi\mu\nu}),
\end{equation}
where $N^{0(+)}_{\rm ST}$ and $N^{0(+)}_{\rm DT}$ are the ST and DT yields in data,
$\epsilon^{0(+)}_{\pi\mu\nu}$ is the signal efficiency of finding $D^{0(+)}\to \pi^{-(0)}\mu^+\nu_\mu$ events in the presence of a ST $\bar D$ meson.
Here, $\epsilon^{0(+)}_{\pi\mu\nu}=\sum_k\frac{N^k_{\rm ST}\epsilon^k_{\rm DT}}{N^{\rm 0(+)}_{\rm ST} \epsilon^k_{\rm ST}}$,
where $N^k_{\rm ST}$ and $\epsilon^k_{\rm ST[DT]}$
are the ST yield and the ST[DT] efficiency of the $k^{\rm th}$ tag mode, respectively.

All charged tracks
are required to be within
a polar-angle range of $|\rm{cos\theta}|<0.93$.
Except for those from $K^0_{S}$ decays,
the good charged tracks
are required to come
from the interaction region defined by
$V_{xy}<$ 1\,cm and $|V_{z}|<$ 10\,cm,
where $V_{xy}$ and $|V_{z}|$
are the distances of closest approach
of the reconstructed track to the interaction point (IP) in
the $xy$ plane and the $z$ direction (along the beam), respectively.
Charged particle identification (PID)
is performed by combining the time-of-flight information with
the specific ionization energy loss measured in the main drift chamber.
The information of the electromagnetic calorimeter (EMC) is also included to identify muon candidates.
Combined confidence levels
for electron, muon, pion and kaon hypotheses ($CL_{e}$, $CL_{\mu}$, $CL_{\pi}$ and $CL_{K}$) are calculated individually. The kaon and pion are required to satisfy $CL_{K}>CL_{\pi}$ and $CL_{\pi}>CL_{K}$, respectively, while muon candidates are selected with $CL_{\mu}> 0.001$, $CL_{\mu}>CL_e$ and $CL_{\mu}>CL_K$. Additionally, muon candidates are required to deposit an energy in the EMC within the range $(0.1,\,0.3)$\,GeV and to satisfy a polar angle and momentum dependent hit depth criterion in the muon counter (MUC)~\cite{pbins}; these criteria suppress the number of pions misidentified as muons.
The $K^0_{S}$ candidate is reconstructed from two oppositely charged tracks with $|V_{z}|< 20$\,cm. These two charged tracks are assumed to be pions (without PID), constrained to a common vertex and are required to have an invariant mass satisfying $|M_{\pi^{+}\pi^{-}} - M_{K_{S}^{0}}|< 12$\,MeV$/c^{2}$, where $M_{K_{S}^{0}}$ is the $K^0_{S}$ nominal mass~\cite{pdg2014}. A selected $K^0_S$ candidate must have a decay length larger than two times of the vertex resolution away from the IP.
Photon candidates are selected from the shower clusters in the EMC that are not associated with a charged track.
The shower time is required to be within 700\,ns of the event start time,
its energy is required to be greater than 25 (50)\,MeV in the EMC barrel (endcap) region~\cite{BESIII}.
The opening angle between the shower and
any charged tracks must be greater than $10^{\circ}$.
A $\pi^0$ candidate is reconstructed from a $\gamma\gamma$ pair with an invariant mass $M_{\gamma\gamma}$ within $(0.115,\,0.150)$\,GeV$/c^{2}$.
A kinematic fit constraining $M_{\gamma\gamma}$ to the $\pi^0$ nominal mass~\cite{pdg2014} is imposed to improve its momentum resolution.

\begin{figure}[htp]
  \centering
  \includegraphics[width=\columnwidth]{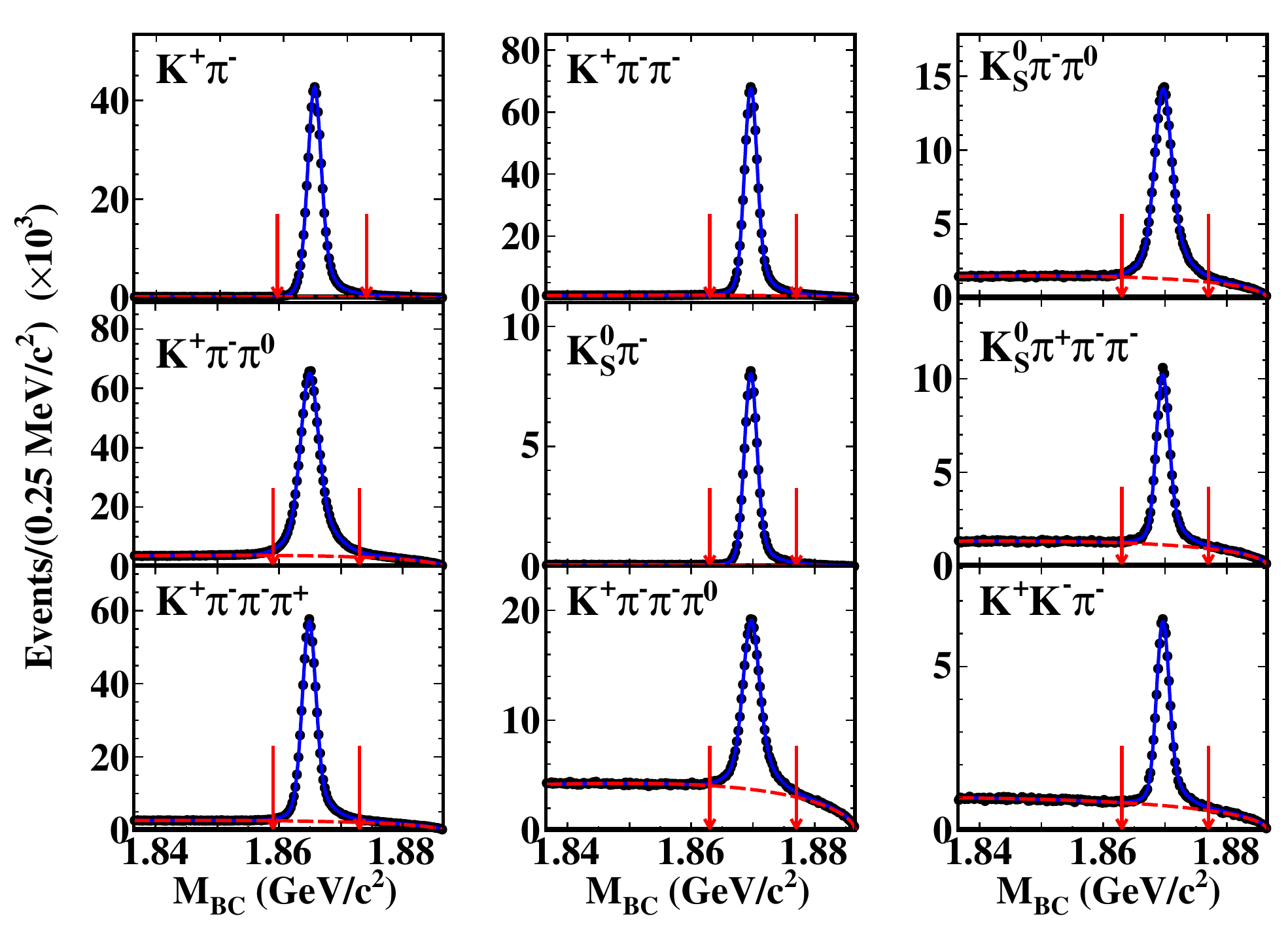}
  \caption{
  (Color online) Fits to the $M_{\rm BC}$ distributions of
  the ST $\bar D^0$ (left column) and $D^-$ (middle and right columns) modes.
The dots with error bars are data.
The blue solid and red dashed curves are the fit results
and the fitted backgrounds.
The signal region is between the red arrows.
}\label{fig:datafit_Massbc}
\end{figure}

The ST $\bar D$ mesons are identified by
the energy difference $\Delta E \equiv E_{\bar D} - E_{\rm beam}$ and
the beam-constrained mass $M_{\rm BC} \equiv \sqrt{E^{2}_{\rm beam}/c^4-|\vec{p}_{\bar D}|^{2}/c^2}$.
Here, $E_{\rm beam}$ is the beam energy,
$\vec{p}_{\bar D}$ and $E_{\bar D}$ are the momentum and energy of the
$\bar D$ candidate in the $e^+e^-$ rest frame.
For each ST mode, if there are multiple candidates in an event,
only the one with the smallest $|\Delta E|$ is kept.
The ST candidates are required to have
$\Delta E\in(-55,\, 40)$~MeV and $(-25,\, 25)$\,MeV for the modes
with and without a $\pi^0$ in the final states, respectively.
For the ST candidates of $\bar D^0\to K^+\pi^-$, the backgrounds from
cosmic rays and Bhabha events are further rejected using the requirements described in Ref.~\cite{deltakpi}.
After the above selection criteria, the
ST yields
are obtained by performing maximum likelihood fits to the $M_{\rm BC}$ distributions for individual ST modes, as shown in Fig.~\ref{fig:datafit_Massbc}.
In the fits, the $\bar D$ signal is modeled by a MC-simulated shape convolved with
a double Gaussian function that describes any resolution difference between data and MC simulation.
For individual tags, the peaks and resolutions of the convolved Gaussian functions
fall in the regions of $(-0.3,0.3)$~MeV/$c^2$ and $(0.7,3.2)$~MeV/$c^2$, respectively.
The combinatorial background is described by an ARGUS function~\cite{ARGUS}.
The candidates in the $M_{\rm BC}$ signal regions, defined as $(1.859,\,1.873)$\,GeV/$c^2$ and $(1.863,\,1.877)$\,GeV/$c^2$ for $\bar D^0$ and $D^-$, respectively,
are kept for further analysis.

In the part of the event recoiling against the ST $\bar D$ meson,
the SL decay candidate is selected from the remaining tracks that have not been used for tag reconstruction.
Events containing a muon candidate, with opposite charge to the ST $\bar D$ candidate, and a $\pi^{-(0)}$ candidate are considered as SL $D^{0(+)}$ decays. We require there are no additional charged tracks in the event.
The potential backgrounds from
$D^{0} \to K^{-}\pi^+$, $D^{0(+)} \to \pi^{-(0)}\pi^+$ and
$D^{0(+)} \to \pi^{-(0)}\pi^+\pi^0/\eta/\bar K^0$ are suppressed by the optimized requirements of
$M_{\pi^{-(0)}\mu^+}<1.7$\,GeV/$c^2$ and $E^{{\rm extra}\,\gamma}_{\rm max}<0.07$\,GeV,
where $M_{\pi^{-(0)}\mu^+}$ is the $\pi^{-(0)}\mu^+$ invariant mass
and $E^{{\rm extra}\,\gamma}_{\rm max}$ is the maximum energy of any additional photon candidates
unused in the DT reconstruction.
The relative efficiencies of the requirements on $M_{\pi^{-(0)}\mu^+}$ and $E^{{\rm extra}\,\gamma}_{\rm max}$
are approximately 99\% and 70\%, respectively.
To further reject the peaking backgrounds of $D^0\to K^0_S(\pi^+\pi^-)\pi^0$ and $D^+\to \bar K^0\pi^+$
for $D^0\to\pi^-\mu^+\nu_\mu$ and $D^+\to\pi^0\mu^+\nu_\mu$,
we require $M_{\pi^{-}\mu^+}$
and $M^{\rm rec}_{D^-\mu^+}$ ($D^-\mu^+$ recoil mass) to be
outside the ranges $(0.46,\,0.50)$\,GeV/$c^2$ and $(0.45,\,0.55)$\,GeV/$c^2$, respectively.
The undetected neutrino is inferred from the variable
$M^2_{\rm miss} \equiv  E^2_{\rm miss}/c^4 - |\vec{p}_{\rm miss}|^2/c^2,$
which peaks at zero for signal events. Here $E_{\rm miss}$ and $|\vec{p}_{\rm miss}|$ are
the missing energy and momentum calculated by
$E_{\rm miss} \equiv E_{\rm beam} - E_{\pi^{-(0)}} - E_{\mu^+}$ and
    $\vec{p}_{\rm miss} \equiv \vec{p}_{D} - \vec{p}_{\pi^{-(0)}} - \vec{p}_{\mu^+},$
in which
$E_{\pi^{-(0)}}\,(E_{\mu^+})$ and $\vec{p}_{\pi^{-(0)}}\,(\vec{p}_{\mu^+})$ are the
energy and momentum of $\pi^{-(0)}$ ($\mu^{+}$) in the rest frame of $e^+e^-$ system.
Furthermore, $\vec{p}_{D} \equiv
(-\hat{p}_{\bar D})
\sqrt{E_{\rm beam}^{2}/c^2-M_{D}^{2}c^2}$ is the momentum of $D$ meson,
where $\hat{p}_{\bar D}$ is the momentum direction
of the ST $\bar D$ meson and $M_{D}$ is the $D$ nominal mass~\cite{pdg2014}.

\begin{figure}[htp]
  \centering
  \includegraphics[width=\columnwidth]{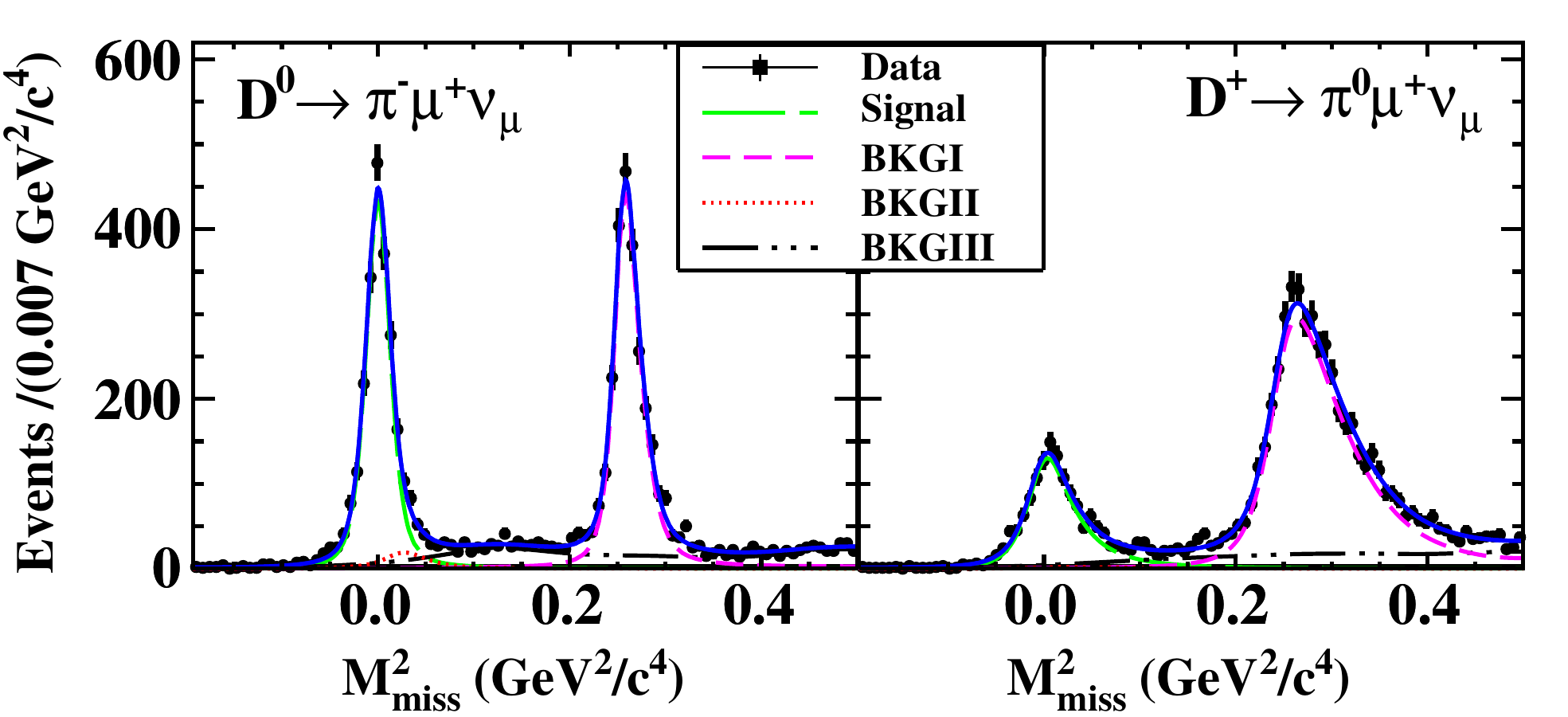}
  \caption{
   (Color online)
  Fits to the $M^2_{\rm miss}$ distributions of the
DT candidates.
The dots with error bars are data.
The blue solid,
green long dashed, pink dashed, red dotted and black dot-dashed curves
represent the overall fit results, the SL signals,
the BKGI, BKGII and BKGIII components (see text), respectively.}
\label{fig:fit_Umistry1}
\end{figure}

Figure~\ref{fig:fit_Umistry1} shows the $M^2_{\rm miss}$
distributions of the selected DT candidates for $D^0\to \pi^-\mu^+\nu_\mu$ and $D^+\to \pi^0\mu^+\nu_\mu$.
Both the candidate events contain two peaks corresponding to the
$D^{0(+)}\to \pi^{-(0)}\mu^+\nu_\mu$ signals and the $D^{0(+)}\to \pi^{-(0)}\pi^+\bar K^0$ backgrounds
(named BKGI) at zero and 0.25 GeV$^2/c^4$, respectively.
MC studies indicate that the small peaking backgrounds from decays $D^0\to K^- \pi^+$, $D^{0(+)}\to \pi^{-(0)} \pi^+$ and
$D^{0(+)}\to \pi^{-(0)} \pi^+\pi^0$ (named BKGII) peak around 0.02\,GeV$^2$/$c^4$, under the right side of signal.
The DT signal yields are determined
by performing unbinned maximum likelihood fits on the $M^2_{\rm miss}$ distributions.
In the fits, the signals,
the peaking backgrounds of BKGI and BKGII
and
other non-peaking backgrounds (named BKGIII)
are described by the corresponding MC-simulated shapes.
The signal, BKGI and BKGII shapes are smeared with
Gaussian functions with free parameters
to take into account the resolution difference between data and MC simulation.
The parameters of the Gaussian function for BKGII are the same as those for the signal, while those for BKGI can be different.
All but one of the BKGII peaking background yields  are fixed to the values from MC simulation; the exception is the  $D^{0}\to \pi^{+}\pi^-\pi^0$ background to the $D^{0}\to \pi^{-}\mu^+\nu_\mu$ signal, which is determined from data due to its good separation from the signal. All the other background component yields are floated in the fit.

The ST and DT yields, the detection efficiencies and the obtained BFs
are shown in Table~\ref{tab:singletagN}.
In BF measurements using the DT method,
the uncertainties from the ST selection mostly cancel.
The relative systematic uncertainties from the different sources considered are shown in Table~\ref{tab:sys}.
The uncertainty from the ST yield is taken as 0.5\%
by examining its relative change between data and MC simulation by
varying the fit range, signal shape and endpoint of ARGUS function.
The efficiencies of $\mu^+$ and $\pi^-$ tracking (PID) and $\pi^0$ reconstruction
are verified using $e^+e^-\to\gamma\mu^+\mu^-$ events and DT $D\bar D$ hadronic events, respectively.
We assign
the uncertainties of $\pi^-$ tracking (PID), $\mu^+$ tracking (PID) and $\pi^0$ reconstruction
to be 0.5\%\,(0.5\%), 0.5\%\,(0.5\%) and 1.0\%, respectively.
The uncertainty related to the choice of the $E_{\rm max}^{\rm extra\,\gamma}$
requirement is assigned by analyzing the control sample $D^{0(+)}\to\pi^{-(0)}e^+\nu_e$; it is 1.2\%\,(1.7\%) for the $D^{0(+)}$ decay.
The uncertainty associated with the $M_{\pi\mu^+}$ requirement
is investigated by using the alternative requirements of 1.65\,GeV/$c^2$ or 1.75\,GeV/$c^2$.
The uncertainty due to the $K^0_S$ veto is estimated by varying the $M_{\pi^-\mu^+}$
($M^{\rm rec}_{D^-\mu^+}$) requirement by $\pm 0.01$\,GeV/$c^2$.
The changes to the measured BFs with the different requirements are taken as the systematic uncertainties.
The uncertainties related to the $M^2_{\rm miss}$ fits are investigated
by varying the fit ranges by $\pm0.025\,(0.050)$\,GeV$^2/c^4$ for $D^{0(+)}$ decays,
and with different parameterizations of signals,
combinatorial and peaking backgrounds.
The effects due to signal shapes are estimated with different requirements on the MC-truth matched signal shapes. The relative magnitudes of the dominant combinatorial background components in BKGIII are varied by $\pm20\%$. The fixed magnitudes of the dominant peaking backgrounds in BKGII are
changed according to the BF uncertainties
\cite{pdg2014}, the limited MC statistics of background channels,
and the data-MC differences of the rates of misidentifying $K^-$ as $\pi^-$
and $\pi^+$ as $\mu^+$.
The maximum changes of BFs
are taken as their respective uncertainties.
The uncertainties due to limited MC statistics are 0.3\% for both decays.
The uncertainty related to MC generator assumptions
is estimated to be 0.3\% via comparing the DT efficiencies
by varying the quoted vector HFF parameters by $\pm 1$ standard deviation 
and replacing the nominal scalar HFF model with the simple pole model~\cite{BK-model}.
The uncertainty due to FSR effect is assigned as 0.3\%,
which is obtained by comparing the nominal DT efficiency to that
when the FSR photon probability is changed by $\pm 20\%$.
The total systematic uncertainty is the quadratic sum of the individual contributions.

\begin{table}[htp]
\centering
\caption{\small
ST and DT yields, signal efficiencies in the $M_{\rm BC}$ signal regions,
and the obtained BFs.
The numbers in the first and second brackets are the statistical and systematic uncertainties
in the last two digits, respectively. The efficiencies do not include ${\mathcal B}_{\pi^0\to \gamma\gamma}$.
See Supplemental Material~\cite{Supplemental} for tag dependent numbers.
}\label{tab:singletagN}
\begin{tabular}{lcccc}
  \hline\hline
Mode & $N^{0(+)}_{{\rm ST}}$\,($\times 10^{4}$) & $N^{0(+)}_{{\rm DT}}$ & $\epsilon^{0(+)}_{\pi\mu\nu}$\,(\%) & ${\mathcal B}_{D\to\pi\mu\nu_{\mu}}$\,(\%) \\  \hline
$\pi^-\mu^+\nu_\mu$&$232.1(02)$&$2265(63)$&$35.82(08)$ & $0.272(08)(06)$\\
$\pi^0\mu^+\nu_\mu$&$152.2(02)$&$1335(42)$&$25.36(07)$ & $0.350(11)(10)$\\
  \hline\hline
\end{tabular}
\end{table}

\begin{table}[htp]
\centering
\caption{\small
Relative systematic uncertainties in BF measurements.}
\label{tab:sys}
\centering
\begin{tabular}{ccc}
  \hline\hline
  Source\,(\%)                  &${\mathcal B}^{0}_{\pi\mu\nu}$ & ${\mathcal B}^{+}_{\pi\mu\nu}$ \\  \hline
  ST yields                                  &0.5&0.5\\
  $\mu^+$ tracking                           &0.5&0.5\\
  $\mu^+$ PID                                &0.5&0.5\\
  $\pi^-$ tracking                           &0.5&-- \\
  $\pi^-$ PID                                &0.5&-- \\
  $\pi^0$ reconstruction                     &-- &1.0\\
  $E^{\rm extra\,\gamma}_{\rm max}$ requirement&1.2&1.7\\
  $M_{\pi\mu^{+}}$ requirement               &0.4&0.9\\
  $K^0_S$ veto                               &-- &0.2\\
  $M_{\rm miss}^2$ fit                       &1.6&1.4\\
  MC statistics                              &0.3&0.3\\
  MC generator                               &0.3&0.3\\
  FSR effect                                 &0.3&0.3\\ \hline
  Total                                      &2.4&2.8\\
  \hline\hline
\end{tabular}
\end{table}

Combining the ${\mathcal B}_{D^{0(+)}\to \pi^{-(0)}\mu^+\nu_\mu}$ measured in this work with previous BESIII measurements~\cite{bes3_piev,bes3_pi0ev}
${\mathcal B}_{D^0\to\pi^- e^+\nu_{e}} = (0.295 \pm 0.004_{\rm stat.}\pm 0.003_{\rm syst.})\%$ and
${\mathcal B}_{D^+\to\pi^0 e^+\nu_{e}} = (0.363 \pm 0.008_{\rm stat.}\pm 0.005_{\rm syst.})\%$,
we obtain
${\mathcal R}^0_{\rm LFU}=0.922\pm 0.030_{\rm stat.}\pm0.022_{\rm syst.}$ and
${\mathcal R}^+_{\rm LFU}=0.964\pm 0.037_{\rm stat.}\pm0.026_{\rm syst.}$.
Here, the systematic uncertainties in ST yields,
$\pi^-$ tracking and PID, and $\pi^0$ reconstruction cancel,
and an additional uncertainty of 0.5\% is included to take into account different FSR effects for electron and muon.
The measured values of ${\mathcal R}_{\rm LFU}^{0(+)}$ coincide with the SM expectation $0.985\pm0.002$~\cite{arxiv0985} within
$1.7\sigma$ $(0.5\sigma)$.

\begin{figure}[htp]
  \centering
  \includegraphics[width=\columnwidth]{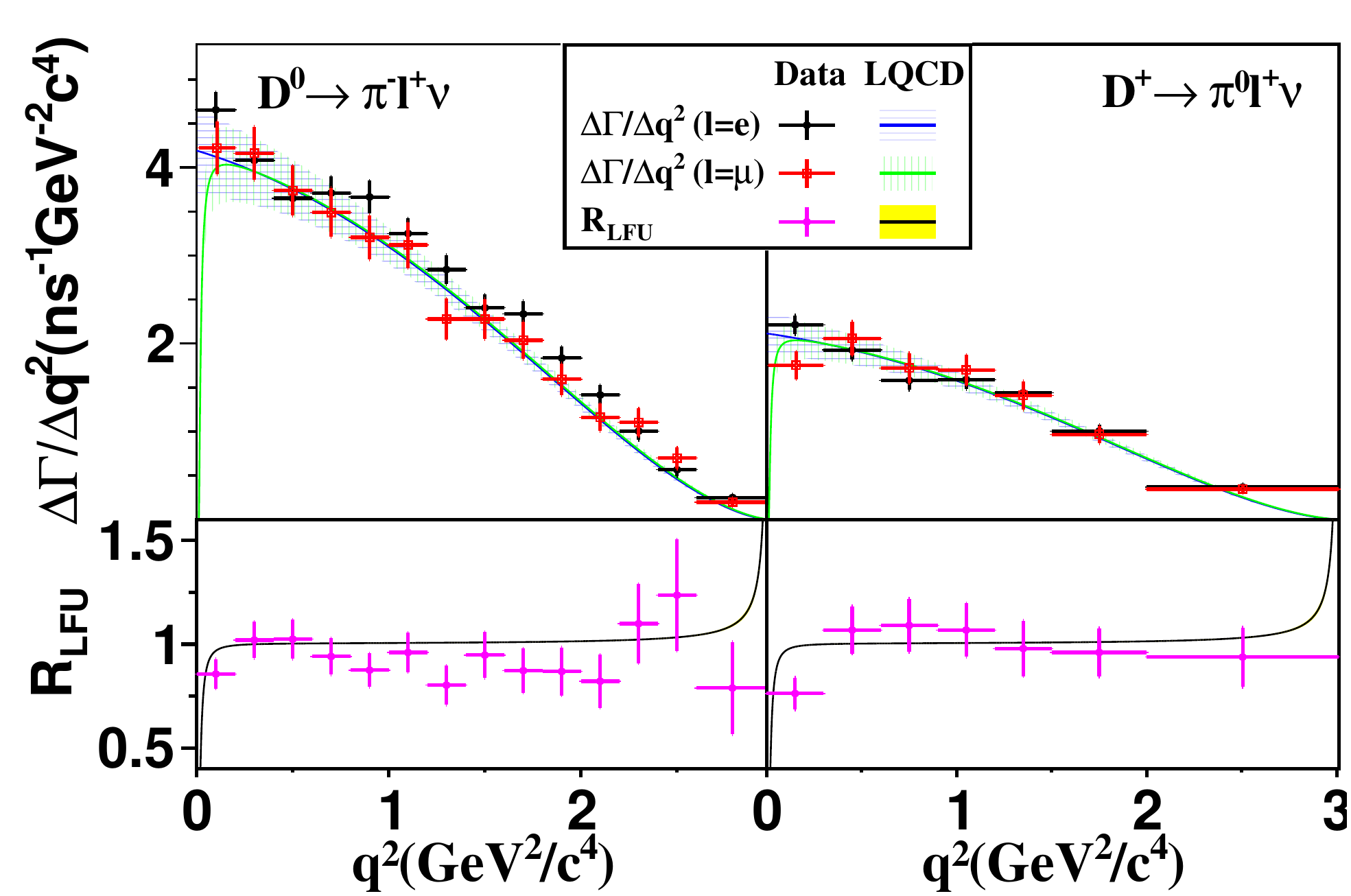}
  \caption{
   (Color online)
$\Delta \Gamma^{0(+)}_{i}/\Delta q^2$ of $D^{0(+)}\to \pi^{-(0)}\ell^+\nu_\ell$ (top)
and ${\mathcal R}^{0(+)}_{\rm LFU}$ (bottom) in various $q^2$ bins.
The calculations of $\Delta \Gamma^{0(+)}_{i}/\Delta q^2$ of $D^{0(+)}\to \pi^{-(0)} e^+\nu_e$ are quoted from
Refs.~\cite{bes3_piev,bes3_pi0ev}.
Data are shown as dots with error bars,
where the uncertainties are combined from statistical and systematic errors, and the uncertainties
in ${\mathcal R}^{0(+)}_{\rm LFU}$ are dominated by the statistical uncertainties of semi-muonic modes.
The blue, green and black curves with bands show the LQCD predictions with uncertainties,
using the equations and HFF parameters described in Refs.~\cite{arxiv0985,Lubicz},
where the theoretical uncertainties in ${\mathcal R}^{0(+)}_{\rm LFU}$ are tiny
due to strong correlation of the form factors.}
\label{fig:Ratio}
\end{figure}

The BFRs ${\mathcal R}_{\rm LFU}^{0(+)}$ are obtained in the full $q^2$
(four-momentum transfer square of $\mu^+\nu_\mu$) region. To investigate the
$q^2$ dependence of ${\mathcal R}_{\rm LFU}^{0(+)}$, we examine BFRs in different $q^2$ ranges.
Using the method described in Refs.~\cite{bes3_piev,bes3_pi0ev},
the partial width of $D^{0(+)}\to \pi^{-(0)}\mu^+\nu_\mu$ in the $i^{\rm th}$ $q^2$ bin
is calculated by
\begin{equation}\label{eq2}
\Delta \Gamma^{0(+)}_{i} = N_{i}^{0(+)}/(\tau_{D^{0(+)}} N^{0(+)}_{\rm ST}),
\end{equation}
where $\tau_{D^{0(+)}}$ is the lifetime of the $D^{0(+)}$ meson, and $N^{0(+)}_{i}$ is the produced DT yield in the $i^{\rm th}$ $q^2$ bin,
calculated by $N_{i}^{0(+)} = \sum_{j}(\epsilon_{0(+)}^{-1})_{ij} M_{j}^{0(+)}$.
Here $M_{j}^{0(+)}$ is the observed DT yield in the $j^{\rm th}$ $q^2$ bin, $\epsilon_{0(+)}$ is the efficiency matrix and
$(\epsilon_{0(+)})_{ij}$ are the elements of a matrix that describes the efficiency and smearing across $q^2$ bins.
See Supplemental Material~\cite{Supplemental} for the observed and produced DT yields,
efficiency matrices as well as the partial widths for $D^{0(+)}\to \pi^{-(0)}\mu^+\nu_\mu$.
Combining with the measured partial widths for $D^{0(+)}\to \pi^{-(0)}e^+\nu_e$ in the same $q^2$
bins~\cite{bes3_piev,bes3_pi0ev}, we obtain ${\mathcal R}^{0(+)}_{\rm LFU}$ in various $q^2$ bins.
Figure~\ref{fig:Ratio} shows $\Delta \Gamma^{0(+)}_{i}/\Delta q^2$ and ${\mathcal R}^{0(+)}_{\rm LFU}$
in various $q^2$ bins, as well as the LQCD predictions for comparison.
The measured values are consistent with the SM predictions within $2\sigma$ in most of the $q^2$ regions.

In summary, using 2.93\,fb$^{-1}$ $e^+e^-$ collision data
collected at $\sqrt{s}=3.773$\,GeV with the BESIII detector,
we have measured the BFs of $D^0\to \pi^-\mu^+\nu_\mu$ and $D^+\to \pi^0\mu^+\nu_\mu$.
The value of ${\mathcal B}_{D^{0}\to \pi^{-}\mu^+\nu_\mu}$ is
consistent with the world-average value~\cite{pdg2014}
and has much improved precision; ${\mathcal B}_{D^{+}\to \pi^{0}\mu^+\nu_\mu}$ is determined for the first time.
Combining the previous BESIII measurements of $D^{0(+)}\to \pi^{-(0)}e^+\nu_e$,
we calculate the $q^2$-integrated and $q^2$-dependent BFRs, and find no
significant evidence of LFU violation.

The BESIII collaboration thanks the staff of BEPCII and the IHEP computing center for their strong support.
Authors thank S. Simula, L. Riggio, G. Salerno and Wei Wang for helpful discussions.
This work is supported in part by National Key Basic Research Program of China under Contract No. 2015CB856700; National Natural Science Foundation of China (NSFC) under Contracts Nos. 11235011, 11305180, 11335008, 11375170, 11425524, 11475123, 11475164, 11475169, 11605196, 11605198, 11625523, 11635010, 11705192; the Chinese Academy of Sciences (CAS) Large-Scale Scientific Facility Program; the CAS Center for Excellence in Particle Physics (CCEPP); Joint Large-Scale Scientific Facility Funds of the NSFC and CAS under Contracts Nos. U1332201, U1532101, U1532102, U1532257, U1532258, U1732263; CAS under Contracts Nos. KJCX2-YW-N29, KJCX2-YW-N45, QYZDJ-SSW-SLH003; 100 Talents Program of CAS; National 1000 Talents Program of China; INPAC and Shanghai Key Laboratory for Particle Physics and Cosmology; German Research Foundation DFG under Contracts Nos. Collaborative Research Center CRC 1044, FOR 2359; Istituto Nazionale di Fisica Nucleare, Italy; Koninklijke Nederlandse Akademie van Wetenschappen (KNAW) under Contract No. 530-4CDP03; Ministry of Development of Turkey under Contract No. DPT2006K-120470; National Science and Technology fund; The Swedish Research Council; U. S. Department of Energy under Contracts Nos. DE-FG02-05ER41374, DE-SC-0010118, DE-SC-0010504, DE-SC-0012069; University of Groningen (RuG) and the Helmholtzzentrum fuer Schwerionenforschung GmbH (GSI), Darmstadt; WCU Program of National Research Foundation of Korea under Contract No. R32-2008-000-10155-0.

\newpage
\onecolumngrid
\renewcommand\thetable{\Roman{table}}
\setcounter{table}{0}
\begin{center}
SUPPLEMENTAL MATERIAL
\end{center}

Table~\ref{tab:singletagN} shows
tag dependent ST yields and efficiencies, DT efficiencies and signal efficiencies of $D^{0(+)}\to \pi^{-(0)}\mu^+\nu_\mu$.
Table~\ref{tab:tab2} and Table~\ref{tab:tab3}
present the efficiency matrices $(\epsilon_{0(+)})_{ij}$,
the range of each $q^2$ bin, the number
of the observed DT events $M_{j}$, the number of produced DT events
$N_{i}$, and the partial decay rate $\Delta\Gamma_i$ in each $q^2$ bin
for $D^{0}\to \pi^{-}\mu^+\nu_\mu$ and $D^{+}\to \pi^{0}\mu^+\nu_\mu$ decays, respectively.

\begin{table}[htp]
\centering
\caption{\small
Summary of
tag dependent ST yields and efficiencies (in \%) in the $M_{\rm BC}$ signal regions,
DT efficiencies and signal efficiencies of $D^{0(+)}\to \pi^{-(0)}\mu^+\nu_\mu$.
The efficiencies do not include ${\mathcal B}_{K^0_S\to \pi^+\pi^-}$ and
   ${\mathcal B}_{\pi^0\to\gamma\gamma}$.
Uncertainties are statistical only.
The variations in $\epsilon^{0(+)}_{\pi\mu\nu,\,k}$
for different ST modes arise mainly from
the $E^{{\rm extra}\,\gamma}_{\rm max}$ requirement.
}\label{tab:singletagN}
\begin{tabular}{lcccc}
  \hline\hline
  ST   mode & $N^{0(+)}_{{\rm ST},\,k}$ & $\epsilon^{0(+)}_{{\rm ST},\,k}$ &   $\epsilon^{0(+)}_{{\rm DT},\,k}$ &
  $\epsilon^{0(+)}_{\pi\mu\nu,\,k}$\\  \hline
$K^+\pi^-$          &  $516971\pm\hspace{0.15cm}746$                   & $64.28\pm0.09$ &$24.15\pm0.05$                &   $37.57\pm0.10$\\
$K^+\pi^-\pi^0$     & $\hspace{-0.15cm}1099361\pm1327$                 & $36.35\pm0.04$ &$14.20\pm0.04$                &   $39.06\pm0.13$\\
$K^+\pi^-\pi^-\pi^+$& $704677\pm1094$                                  & $37.84\pm0.05$ &$11.15\pm0.06$                &   $29.46\pm0.16$\\ \hline
$K^+\pi^-\pi^-$        & $782669\pm\hspace{0.15cm}990$                 & $50.57\pm0.06$ &$12.34\pm0.04$                &   $24.41\pm0.09$\\
$K^0_S\pi^-$           & $\hspace{0.15cm}91345\pm\hspace{0.15cm}320$   & $50.39\pm0.17$ &$14.41\pm0.05$                &   $28.60\pm0.14$\\
$K^+\pi^-\pi^-\pi^0$   & $251008\pm1135$                               & $26.72\pm0.09$ &$\hspace{0.15cm}6.32\pm0.04$  &   $23.66\pm0.16$\\
$K^0_S\pi^-\pi^0$      & $215364\pm1238$                               & $27.25\pm0.07$ &$\hspace{0.15cm}8.33\pm0.06$  &   $30.58\pm0.24$\\
$K^0_S\pi^+\pi^-\pi^-$ & $113054\pm\hspace{0.15cm}889$                 & $28.29\pm0.12$ &$\hspace{0.15cm}7.56\pm0.06$  &   $26.72\pm0.25$\\
$K^+K^-\pi^-$          & $\hspace{0.15cm}69034\pm\hspace{0.15cm}460$   & $40.87\pm0.24$ &$\hspace{0.15cm}7.99\pm0.05$  &   $19.55\pm0.16$\\
  \hline\hline
\end{tabular}
\end{table}

\begin{table*}[htp]
\centering
\caption{
Summary of the efficiency matrix $(\epsilon_{0(+)})_{ij}$  (in \%),
the range of each $q^2$ bin (in GeV$^2$/$c^4$), the number
of the observed DT events $M_{j}$, the number of produced DT events
$N_{i}$ and the partial decay rate $\Delta\Gamma_i$ (in ns$^{-1}$) in each $q^2$ bin
for $D^{0}\to \pi^{-}\mu^+\nu_\mu$.
The column of $(\epsilon_{0(+)})_{ij}$ gives the true $q^2$ bin $j$, while the row gives the reconstructed $q^2$ bin $i$.
Uncertainties are statistical only.
For $\Delta\Gamma_i$, the uncertainty is combined from the statistical and systematic uncertainties,
in which the statistical uncertainty for muon mode is dominated.
}\label{tab:tab2}
\resizebox{!}{2.80cm}{
\begin{tabular}{lcccccccccccccc|cccc}
\hline
\hline
$\epsilon_{ij}$&    1&    2&    3&    4&    5&    6&    7&    8&    9&   10&   11&   12&   13&   14&$q^2$ bin&$M_{j}$&$N_{i}$&$\Delta\Gamma_i$\\\hline
 1      &33.56& 0.80& 0.01& 0.00& 0.00& 0.00& 0.00& 0.00& 0.00& 0.00& 0.00& 0.00& 0.00& 0.00&($m_\mu^2$,\,0.2)           &261$\pm$17&759$\pm$51&0.797$\pm$0.057\\
 2      & 0.94&33.68& 1.11& 0.02& 0.01& 0.00& 0.00& 0.00& 0.00& 0.00& 0.00& 0.00& 0.00& 0.00&(0.2,\,0.4)           &282$\pm$18&792$\pm$54&0.832$\pm$0.060\\
 3      & 0.04& 1.23&35.25& 1.17& 0.03& 0.01& 0.00& 0.00& 0.00& 0.00& 0.00& 0.01& 0.01& 0.00&(0.4,\,0.6)           &269$\pm$18&712$\pm$51&0.748$\pm$0.057\\
 4      & 0.02& 0.08& 1.31&36.01& 1.20& 0.03& 0.01& 0.00& 0.00& 0.00& 0.01& 0.00& 0.00& 0.00&(0.6,\,0.8)           &257$\pm$18&664$\pm$50&0.698$\pm$0.055\\
 5      & 0.02& 0.05& 0.07& 1.42&35.52& 1.29& 0.03& 0.00& 0.00& 0.00& 0.00& 0.00& 0.00& 0.00&(0.8,\,1.0)           &235$\pm$16&610$\pm$45&0.641$\pm$0.050\\
 6      & 0.01& 0.03& 0.05& 0.12& 1.37&33.87& 1.12& 0.01& 0.00& 0.01& 0.00& 0.00& 0.00& 0.00&(1.0,\,1.2)           &216$\pm$16&594$\pm$47&0.624$\pm$0.052\\
 7      & 0.01& 0.03& 0.02& 0.06& 0.08& 1.41&32.84& 1.07& 0.03& 0.01& 0.00& 0.00& 0.00& 0.00&(1.2,\,1.4)           &157$\pm$14&434$\pm$43&0.456$\pm$0.046\\
 8      & 0.01& 0.01& 0.02& 0.04& 0.05& 0.12& 1.44&31.20& 0.95& 0.06& 0.00& 0.00& 0.00& 0.00&(1.4,\,1.6)           &147$\pm$13&434$\pm$42&0.456$\pm$0.045\\
 9      & 0.00& 0.01& 0.02& 0.02& 0.03& 0.06& 0.12& 1.16&30.72& 1.10& 0.03& 0.01& 0.00& 0.00&(1.6,\,1.8)           &129$\pm$12&388$\pm$39&0.407$\pm$0.043\\
10      & 0.00& 0.00& 0.00& 0.01& 0.04& 0.04& 0.08& 0.15& 1.19&29.91& 0.97& 0.01& 0.00& 0.00&(1.8,\,2.0)           & 99$\pm$10&303$\pm$34&0.319$\pm$0.036\\
11      & 0.00& 0.00& 0.00& 0.00& 0.00& 0.01& 0.03& 0.07& 0.18& 1.34&30.97& 1.02& 0.04& 0.00&(2.0,\,2.2)           & 76$\pm$9 &221$\pm$29&0.232$\pm$0.031\\
12      & 0.00& 0.00& 0.00& 0.00& 0.00& 0.00& 0.00& 0.02& 0.06& 0.12& 1.23&30.58& 0.94& 0.02&(2.2,\,2.4)           & 69$\pm$9 &210$\pm$30&0.221$\pm$0.032\\
13      & 0.00& 0.00& 0.00& 0.00& 0.00& 0.00& 0.00& 0.00& 0.01& 0.05& 0.20& 1.17&31.16& 0.65&(2.4,\,2.6)           & 45$\pm$7 &133$\pm$23&0.140$\pm$0.024\\
14      & 0.00& 0.00& 0.00& 0.00& 0.00& 0.00& 0.00& 0.00& 0.00& 0.00& 0.03& 0.09& 1.24&29.90&(2.6, $q^2_{\rm max}$)& 23$\pm$5 & 71$\pm$17&0.074$\pm$0.018\\
\hline
\hline
\end{tabular}
}
\end{table*}

\begin{table*}[htp]
\centering
\caption{
Summary of the efficiency matrix $(\epsilon_{0(+)})_{ij}$  (in \%),
the range of each $q^2$ bin (in GeV$^2$/$c^4$), the number
of the observed DT events $M_{j}$, the number of produced DT events
$N_{i}$ and the partial decay rate $\Delta\Gamma_i$ (in ns$^{-1}$) in each $q^2$ bin
for $D^{+}\to \pi^{0}\mu^+\nu_\mu$.
The column of $(\epsilon_{0(+)})_{ij}$ gives the true $q^2$ bin $j$, while the row gives the reconstructed $q^2$ bin $i$.
Uncertainties are statistical only.
For $\Delta\Gamma_i$, the uncertainty is combined from the statistical and systematic uncertainties,
in which the statistical uncertainty for muon mode is dominated.
}\label{tab:tab3}
\begin{tabular}{lccccccc|cccc}
  \hline\hline
$\epsilon_{ij}$ &     1 &     2 &     3 &     4 &     5 &     6 &     7 &$q^2$ bin&$M_{j}$&$N_{i}$&$\Delta\Gamma_i$\\\hline
 1              & 26.58 &  0.46 &  0.00 &  0.00 &  0.00 &  0.00 &  0.00 &($m_\mu^2$,\,0.3)            & 215$\pm$18 &    792$\pm$68&0.506$\pm$0.046\\
 2              &  2.13 & 25.16 &  0.62 &  0.01 &  0.00 &  0.00 &  0.00 &(0.3,\,0.6)            & 265$\pm$21 &    966$\pm$84&0.617$\pm$0.057\\
 3              &  0.04 &  2.79 & 24.17 &  0.65 &  0.00 &  0.00 &  0.00 &(0.6,\,0.9)            & 228$\pm$18 &    809$\pm$76&0.517$\pm$0.051\\
 4              &  0.01 &  0.05 &  2.94 & 22.16 &  0.60 &  0.01 &  0.01 &(0.9,\,1.2)            & 205$\pm$17 &    797$\pm$78&0.509$\pm$0.052\\
 5              &  0.01 &  0.02 &  0.05 &  2.74 & 19.32 &  0.35 &  0.00 &(1.2,\,1.5)            & 153$\pm$14 &    662$\pm$74&0.423$\pm$0.049\\
 6              &  0.01 &  0.02 &  0.04 &  0.08 &  2.40 & 18.96 &  0.29 &(1.5,\,2.0)            & 162$\pm$14 &    756$\pm$75&0.483$\pm$0.050\\
 7              &  0.03 &  0.04 &  0.06 &  0.10 &  0.19 &  1.46 & 19.94 &(2.0,\,$q^2_{\rm max}$)& 123$\pm$13 &    546$\pm$65&0.349$\pm$0.043\\
  \hline\hline
\end{tabular}
\end{table*}


\begin{thebibliography}{**}
\bibitem{charge}
Throughout the Letter, the charge conjugate processes are implied unless explicitly stated otherwise.
\bibitem{pbins}
Given six momenta $a$, $b$, $c$, $d$, $e$, and $f$,
the five momenta ranges are defined as
$(a,\,b)$,
$(b,\,c)$,
$(c,\,d)$,
$(d,\,e)$ and
$(e,\,f)$, respectively.
Two dimensional
($|\cos\theta_\mu|$ and momentum $p_\mu$) requirement on
the hit depth ($\mathcal D$) in the MUC is imposed
if $p_\mu>0.5$\,GeV/$c$~\cite{BESIII}.
The $|\cos\theta_{\mu,\,i}|$
is equally divided as
$[0.2\times (i-1),\,0.2\times i)~(i=1,\, 2, \, 3,\, 4,\, {\rm or}~5)$.
In the first three $|\cos\theta_{\mu,\,i}|$ bins,
we require $\mathcal D$ greater than
3, $100\times p_\mu-(55+3\times i)$, 17, $100\times p_\mu-(68+3\times i)$ and
33\,cm
for $p_\mu$ in the ranges bounded by
every neighbour two momenta of $0.50,\,0.58+0.03\times i,\,0.72+0.03\times i,
\,0.85+0.03\times i,\,1.01+0.03\times i$ and 1.2~\cite{pbins}, respectively.
For the other two $|\cos\theta_{\mu,\,i}|$ bins,
we require $\mathcal D$ greater than 9\,cm uniformly.
\end{thebibliography}
\end{document}